\documentclass[letterpaper, 10 pt, conference]{ieeeconf}  
\pdfoutput=1 

\IEEEoverridecommandlockouts                              

\usepackage{graphics} 
\usepackage{graphicx}
\usepackage{multirow}

\title{\LARGE \bf
Pattern Recognition in Vital Signs Using Spectrograms
}

\author{Sidharth Srivatsav Sribhashyam$^{1}$, Md Sirajus Salekin$^{1}$, Dmitry Goldgof$^{1}$, Ghada Zamzmi$^{1}$, Mark Last$^{2}$, and \\Yu Sun$^{1}$ 
\thanks{$^{1}$Department of Computer Science and Engineering, University of South Florida, Tampa, Florida, United States}%
\thanks{$^{2}$Department of Software and Information Systems Engineering, Ben-Gurion University of the Negev, Israel}%
\thanks{*This research is partially supported by University of South Florida Nexus Initiative (UNI) Grant and National Institutes of Health Grant (NIH R21NR018756).}
}

\usepackage{tikz}
\newcommand\copyrighttext{%
  \footnotesize \textcopyright \hspace{0.01cm} 2021 IEEE.  Personal use of this material is permitted.  Permission from IEEE must be obtained for all other uses, in any current or future media, including reprinting/republishing this material for advertising or promotional purposes, creating new collective works, for resale or redistribution to servers or lists, or reuse of any copyrighted component of this work in other works.
  }
\newcommand\copyrightnotice{%
\begin{tikzpicture}[remember picture,overlay]
\node[anchor=south,yshift=10pt] at (current page.south) {\fbox{\parbox{\dimexpr\textwidth-\fboxsep-\fboxrule\relax}{\copyrighttext}}};
\end{tikzpicture}%
}

\begin{document}
\maketitle
\copyrightnotice
\thispagestyle{empty}
\pagestyle{plain}

\begin{abstract}
Spectrograms visualize the frequency components of a given signal which may be an audio signal or even a time-series signal. Audio signals have higher sampling rate and high variability of frequency with time. Spectrograms can capture such variations well. But, vital signs which are time-series signals have less sampling frequency and low-frequency variability due to which, spectrograms fail to express variations and patterns. In this paper, we propose a novel solution to introduce frequency variability using frequency modulation on vital signs. Then we apply spectrograms on frequency modulated signals to capture the patterns. The proposed approach has been evaluated on 4 different medical datasets across both prediction and classification tasks. Significant results are found showing the efficacy of the approach for vital sign signals. The results from the proposed approach are promising with an accuracy of 91.55\% and 91.67\% in prediction and classification tasks respectively.


\end{abstract}


\section{INTRODUCTION}

Analyzing the patterns from vital signs can help us to detect or predict physiological abnormalities. In the case of patients in critical care units, continuous monitoring and analysis of vital signs are critical to detect or predict any medical emergency \cite{MarkLast}. For example, a hypotensive \cite{sahni2016hypotension} episode can be defined as an episode of abnormally low blood pressure that may hamper the supply of oxygen and vital nutrients to organs and may cause organ failure which may be fatal \cite{PATEL201076, zenati2002brief}. 

Several works \cite{MarkLast,MOGHADAM2021104120} show that, by analyzing patterns in Blood Pressure continuously, it is possible to predict a hypotensive episode well ahead of time. In \cite{7900284}, a multimodal approach which includes vital signs as one of the parameters is used for detecting pain in neonates. The method combines predictions from different types of data such as facial expression, crying sound, body movements, and vital signs. This work proves that pain can be detected by analyzing vital signs. Vital signs also give information about the stress levels of a person apart from medical conditions. For example, several works \cite{6567512,8361691} have shown that analysis of patterns from vital signs can help us to detect if a person is under mental stress. Different machine learning algorithms have been used to recognize patterns from vital signs. For example, in \cite{MarkLast}, XGBoost algorithm is used to classify hypotensive episodes from normal Blood Pressure. In \cite{MOGHADAM2021104120}, Logistic Regression and Support Vector Machine are used for binary classification of hypotensive episodes. In \cite{wang2015encoding}, vital sign signals (time-series data) are encoded as pixels of an image at first and then analyzed using Gramian angular fields. Spectrograms have been used to represent ECG signals as images. For example, in \cite{898633}, spectrogram has been used for analyzing ECG signals to detect obstructive sleep apnea. There is some research involving spectrogram analysis of Radar signals for detection of vital signs like \cite{7166886, 8676362}. To the best of our knowledge, there is no work analyzing patterns from vital signs specifically Heart Rate, Respiratory Rate, Blood Pressure using spectrograms. This paper explores the usage of spectrograms for pattern recognition from vital signs.

Vital signs have low variability, low sampling frequency, and almost no seasonality. Due to this, it is difficult to extract patterns from vital signs using spectrograms directly. To solve this issue, we propose a novel solution for this using \textit{Frequency Modulation}. Frequency modulation translates the amplitude of a vital signs signal to the frequency of another signal. In essence, the frequency of the other signal carries the information about the amplitude of the vital signs signal. By this method, we introduce high-frequency variability into the signal and also carry information about the amplitude of vital signs in frequency which can be analyzed using spectrograms.

\section{Background}

\subsection{Spectrogram}
\begin{figure}
\includegraphics[width = 0.45\textwidth]{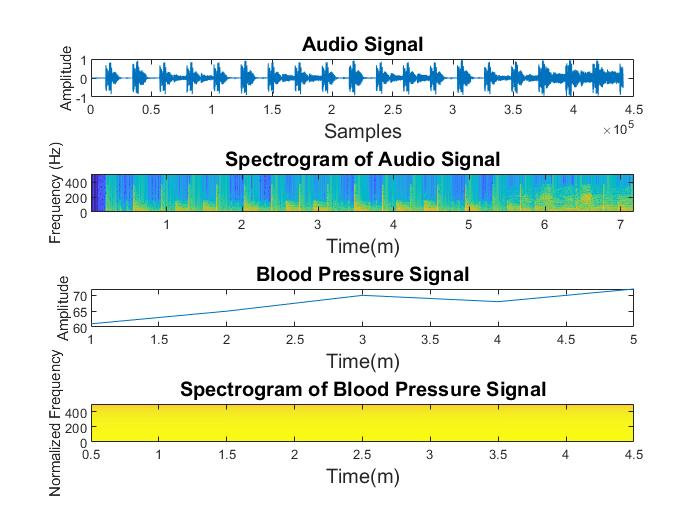}
\centering
  \caption{Examples of audio signal and vital sign signal with their corresponding spectrograms from PIC \cite{PIC} dataset.}
  \label{fig:Spectrogram}
\end{figure}

Spectrograms \cite{oppenheim1999discrete, 8962827} are the visual representations of frequency components of a signal. Spectrogram uses \textit{Fourier Transform} \cite{1} to translate a time-domain signal into frequency domain signal. It divides the whole signal into individual portions or windows and calculates Fourier transform on each window and stitch them back into a single image revealing dominant frequencies in different windows. Spectrograms are typically applied to audio signals as they are best suited for frequency domain analysis. They can also be applied to time-series signals which exhibit seasonality. In Figure \ref{fig:Spectrogram}, we can observe that, as there are high variations in frequency and amplitude in the audio signal and many different frequency components with high sampling frequency, we can easily obtain frequency domain information using spectrograms. On the contrary, in the case of vital sign signal, there is no much variability, sampling frequency is low, and the spectrograms does not exhibit any useful patterns.

\subsection{Frequency Modulation}
Frequency Modulation (FM) \cite{bjorklund1983frequency} is translating information or amplitude of a signal into the frequency of another signal. FM deals with 3 signals namely carrier signal, modulating signal, modulated signal. The carrier signal is typically a high-frequency signal which will be modulated. The modulating signal is a message signal which carries information in its amplitude. Using FM, the information in the message signal is encoded as a proportional frequency of the carrier signal. In essence, the frequency of carrier signal varies proportionally to the amplitude of the message signal thereby carrying the information in its frequency. A frequency-domain analysis on FM signal will reveal the information of the original message signal. Equation for frequency modulation \cite{bjorklund1983frequency} can be given as, $f(t) = A_c cos(2\pi f_ct+m_isin(2\pi f_mt))$
where, $f(t)$ stands for frequency modulated signal in time-domain, $A_c$ stands for carrier signal amplitude, $A_m$ stands for amplitude of message or modulating signal, $t$ stands for instantaneous time, and $m_i$ stands for \textit{modulation index} which decides by how much factor does the frequency of carrier wave varies with the amplitude of modulating signal. Modulation index can be given by $\Delta f/f_m$ where, $\Delta$ f stands for frequency deviation which gives the information about how much frequency should carrier wave change in accordance with the amplitude of modulating signal.

\subsection{Deep Neural Networks and Data Augmentation}
\begin{figure}
\includegraphics[width = 0.45\textwidth]{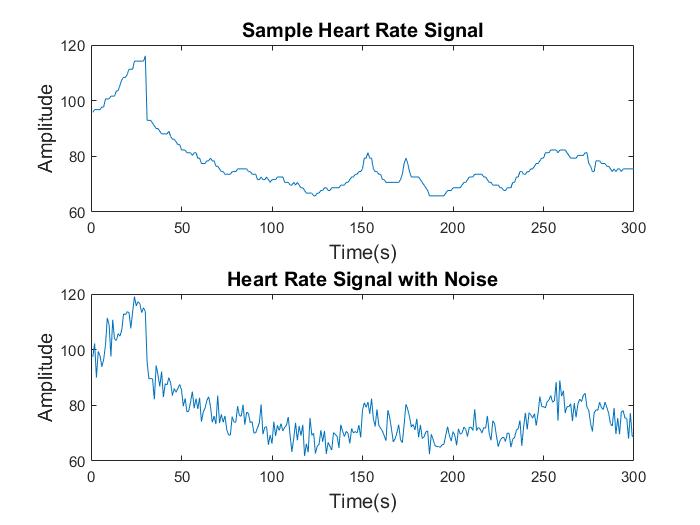}
\centering
  \caption{A Heart Rate signal before and after adding Gaussian noise of Mean = 0 and STD = 3 from NEBD \cite{7780081} dataset.}
  \label{fig:Noise}
\end{figure}

Deep Neural Networks are very popular. Among them, Convolutional Neural Networks (CNN) \cite{salekin2019multi} are used for feature extraction from images. Few popular CNN architectures include VGG16 \cite{simonyan2014very, 8962827} and ResNet \cite{he2016deep, 8962827}. CNNs require large amount of data point which is rare in medical datasets \cite{MIMIC,PIC,7780081,salekin2021106796MultimodalDataset}. To increase the number of samples, data augmentation techniques are performed as described in \cite{8962827, salekin2021multimodal}. We particularly use Gaussian noise which is a set of random numbers distributed normally over a given mean and within a standard deviation. An example of this noise addition is shown in Figure \ref{fig:Noise}. It can be observed that the signal after adding noise well preserves the characteristics of the original signal. 

\section{Methodology}
\begin{figure}
\includegraphics[width = 0.45\textwidth]{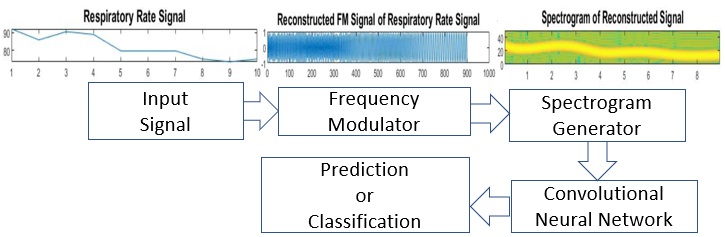}
\centering
\caption{Overall pipeline of the proposed approach.}
 \label{fig:flow}
\end{figure}

The proposed approach has two main steps. The first one
is to reconstruct the vital sign signal using the FM technique and generate the spectrogram from the new modified signals. The second one is to extract the features from the spectrogram using CNN. Figure \ref{fig:flow} shows the overall pipeline of the proposed approach.

\subsection{Reconstruction of Vital Sign Signal Using FM}
\begin{figure}
\includegraphics[width = 0.45\textwidth]{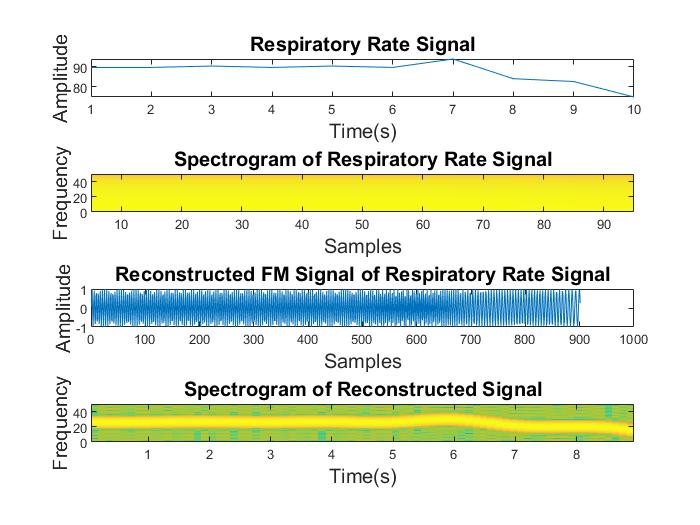}
\centering
\caption{Different representations of Respiratory Rate signal from USF-MNPAD-I \cite{7900284} dataset for No-pain class.}
 \label{fig:pain0}
\end{figure}

As mentioned earlier, vital signs have a low sampling rate, low variability which may not be suited to be analyzed using spectrograms as shown in Figure \ref{fig:pain0}. As vital signs signal is a time-series signal, we need to analyze the instantaneous values of the signal at different points of time. As frequency modulation encodes amplitude as the frequency of another signal, frequency domain analysis of the new signal will provide us information about the original signal. For the purpose of the experiments presented in this paper, the carrier signal frequency($f_c$) was set at $50KHz$ and the frequency deviation ($\Delta f$) used was $850Hz$.

\begin{figure}
\includegraphics[width = 0.45\textwidth]{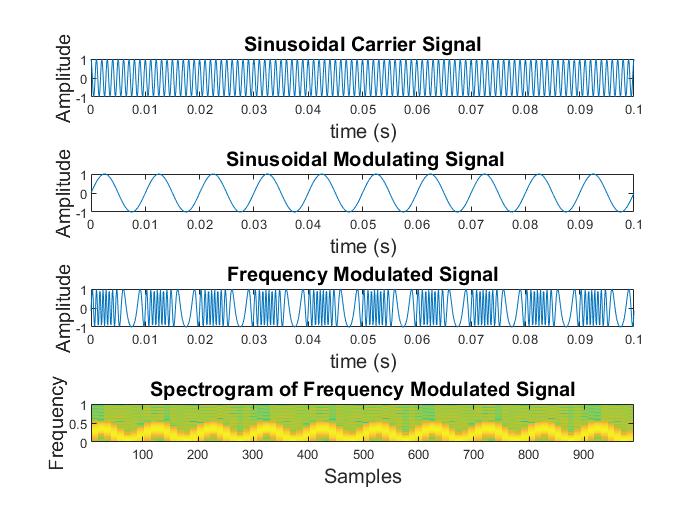}
\centering
\caption{Spectrogram of frequency modulated signal of a random sinusoidal signal.}
 \label{fig:fmrec}
\end{figure}

\begin{figure}
\includegraphics[width = 0.45\textwidth]{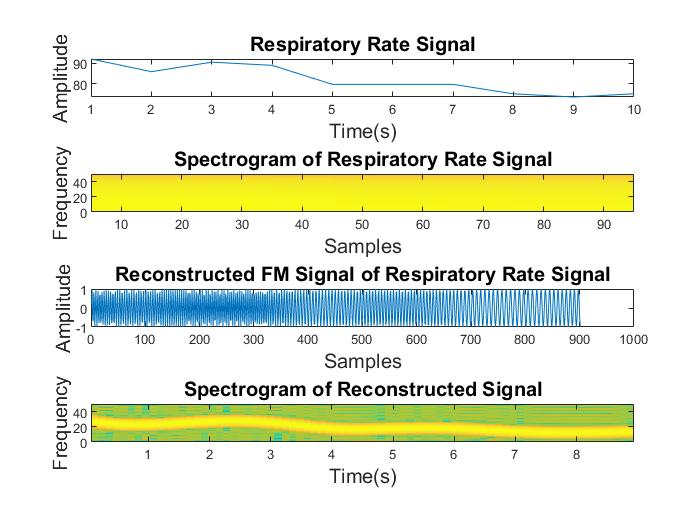}
\centering
\caption{Different representations of Respiratory Rate signal from USF-MNPAD-I \cite{7900284}  dataset for Pain class.}
 \label{fig:pain1}
\end{figure}

Figure \ref{fig:fmrec} shows how the frequency of frequency modulated wave varies proportionally to the amplitude of sinusoidal modulating signal. Applying spectrogram on frequency modulated signal will help us get back the modulating sinusoidal signal as shown in the fourth tile of the Figure. In the Figure \ref{fig:pain0} and Figure \ref{fig:pain1} the vital signs signals, the corresponding reconstructed FM signals, and spectrogram of reconstructed signals are plotted for 2 different classes from USF-MNPAD-I dataset \cite{salekin2021106796MultimodalDataset}. We can see that the frequency of reconstructed signal varies proportionally to the amplitude of vital signs signal. It can also be observed that information from the original vital signs signal is encoded as frequency and that information can be retrieved back by applying spectrogram on the reconstructed signal. But the spectrogram of original signal does not show any patterns comparatively. Spectrograms uncover the frequency components in a given time-series signal and the FM signal's frequency has information of the amplitude of input signal.

\subsection{Classification Using CNN}
The generated spectrograms are fed into CNNs for pattern recognition. VGG16 \cite{simonyan2014very} network with an input image size of $128\times 128$ has been predominantly used in this paper for the classification of spectrograms. Subject-wise cross-validation has been used to validate the model on different portions of data to get a true estimate of the performance of the model. In this experiment, we use \textit{Snapshot Ensemble} \cite{huang2017snapshot} which takes multiple snapshots of weights of models at different epochs during training. Doing this will reduce the consumption of computational resources. For the final predictions, an odd number of snapshots are ensembled by voting method \cite{10.5555/648054.743935}. Further details can be found in the individual experiment sections.

\section{Experimental Results and Discussion}

\subsection{Datasets}
To evaluate the efficacy of the proposed approach, the method is tested on 4 medical datasets. These are MIMIC-III dataset \cite{MIMIC,MIMICIII,goldberger2000physiobank}, Pediatric Intensive Care dataset (PIC) \cite{PIC,goldberger2000physiobank}, USF-MNPAD-I \cite{salekin2021106796MultimodalDataset}, and Non-EEG dataset \cite{7780081,goldberger2000physiobank}. Among these, MIMIC-III \cite{MIMICIII} and PIC \cite{PIC} are used for prediction tasks, while USF-MNPAD-I \cite{salekin2021106796MultimodalDataset} and Non-EEG \cite{7780081} datasets are used for detection tasks.

\subsubsection{MIMIC-III Dataset}
MIMIC-III \cite{MIMIC,MIMICIII,goldberger2000physiobank} consists of deidentified medical data of over 40,000 patients who stayed in critical care units of the Beth Israel Deaconess Medical Center between 2001 and 2012. It includes information such as demographics, vital sign measurements made at the bedside (1 data point per hour), mortality, medications, and so on. The data covers 38,597 distinct adult patients and 49,785 hospital admissions. The median age of adult patients is 65.8 years (Q1–Q3:  52.8–77.8). In this work, this dataset is used to predict an imminent \textit{Hypotensive episode} using Mean Arterial Pressure as the feature obtained from the dataset.

\subsubsection{Paediatric Intensive Care (PIC) Dataset}
The Paediatric Intensive Care (PIC) dataset \cite{PIC,goldberger2000physiobank} consists of medical information of de-identified children admitted to critical care units at a children’s hospital in China. It includes vital sign measurements, medications, laboratory measurements, fluid balance recordings, and so on. The total number of patients in this dataset is 13,941. There is a separate section for vital signs measurement during surgery of neonates which has been used to predict a hypotensive episode. It has information about the specific age category of pediatric patients. The age of subjects ranges from 0-18 years who fall under this category. For the purpose of experiment, only neonates are considered. In this work, we have used systolic pressure obtained during surgery as a feature to predict an imminent hypotensive episode.

\subsubsection{USF-MNPAD-I}
USF-MNPAD-I \cite{salekin2021106796MultimodalDataset} dataset contains 58 neonates which were collected during procedural (acute) and postoperative (prolong acute) procedures. Among them, 13 procedural subjects have vital signs sequences which include Heart Rate, Blood Pressure, and Oxygen saturation. It has already been proved in \cite{7900284} that vital signs give away patterns regarding the pain state of a neonate. In this experiment, similar to \cite{7900284}, classification of pain and no-pain based on vital signs is explored using the reconstructed signal spectrogram approach. As oxygen saturation does not change much and even reconstructed signals cannot give out much patterns, it has been discarded for this experiment. Every sequence has 10 seconds sample with a data point for every 1 second along with ground truth values indicating if the baby is experiencing pain or no-pain.

\subsubsection{A Non-EEG Biosignals Dataset (NEBD)}
Non-EEG Biosignals Dataset for Assessment and Visualization of Neurological Status (NEBD) \cite{7780081,goldberger2000physiobank} contains Non-EEG signals like Heart Rate, SpO2, EDA, Temperature, accelerometer readings of 20 healthy subjects undergoing phases of relaxation and stress for every 5 minutes alternatively. Different signals are sampled at different frequencies. In this work, we use the Heart Rate signal as a feature to detect and classify stress and relaxation.


\begin{table*}[t]
\caption{Performance of different approaches using different metrics}
\label{mimic}
\begin{center}

\begin{tabular}{|c|c|c|c|c|c|c|c|}
\hline

Dataset & Approach & Class & Precision & Recall & F1-score & Accuracy & AUC \\
\hline

\multirow{6}{*}{MIMIC-III} 

& \multirow{3}{*}{Moghadam et al. \cite{9175451}} & Non-hypotensive & 90.55 & \textbf{92.18} & 91.32 & \multirow{3}{*}{90.67}  & \multirow{3}{*}{\textbf{0.96}}\\ \cline{3-6}
& & Hypotensive & 88.00 & 85.00 & 86.33 & & \\ \cline{3-6}
& & Overall & 90.58 & 90.67 & 90.54 & &  \\ \cline{2-8}

& \multirow{3}{*}{Proposed} & Non-hypotensive & \textbf{91.43} & 91.43 & \textbf{91.43} & \multirow{3}{*}{\textbf{91.55}} & \multirow{3}{*}{0.92} \\ \cline{3-6}
& & Hypotensive & \textbf{91.67} & \textbf{91.67}& \textbf{91.67} & - & - \\ \cline{3-6}
& & Overall & \textbf{91.55} & \textbf{91.55} & \textbf{91.55}  &  & \\ \cline{3-8}
\hline

\multirow{6}{*}{PIC} 

& \multirow{3}{*}{Moghadam et al. \cite{9175451}} & Non-hypotensive & 86.90 & 88.24 & 87.36 & \multirow{3}{*}{86.90} & \multirow{3}{*}{\textbf{0.95}} \\ \cline{3-6}
& & Hypotensive & \textbf{86.96} & 85.65 & \textbf{86.08} &  &  \\ \cline{3-6}
& & Overall & 87.32 & 86.90 & 86.90 & & \\ \cline{2-8}

& \multirow{3}{*}{Proposed} & Non-hypotensive & \textbf{88.71} & \textbf{88.99} & \textbf{88.85} & \multirow{3}{*}{\textbf{87.61}} & \multirow{3}{*}{0.87}\\ \cline{3-6}
& & Hypotensive & 86.22 & \textbf{85.88} & 86.05 &  &  \\ \cline{3-6}
& & Overall & \textbf{87.60} & \textbf{87.61} & \textbf{87.61}  & &  \\ \cline{2-8}
\hline

\multirow{6}{*}{NEBD} 

& \multirow{3}{*}{Jafari et al. \cite{8419763}} & Relaxation & 75.00 & 80.00 & 77.42 & \multirow{3}{*}{76.67} & \multirow{3}{*}{0.77} \\ \cline{3-6}
& & Stress & 78.57 & 73.33 & 75.86 &  &  \\ \cline{3-6}
& & Overall & 76.79 & 76.67 & 76.64 & &   \\ \cline{2-8}

& \multirow{3}{*}{Proposed} & Relaxation & \textbf{91.67} & \textbf{91.67} & \textbf{91.67} & \multirow{3}{*}{\textbf{91.67}} & \multirow{3}{*}{\textbf{0.92}} \\ \cline{3-6}
& & Stress & \textbf{91.67} & \textbf{91.67} & \textbf{91.67} &  &  \\ \cline{3-6}
& & Overall & \textbf{91.67} & \textbf{91.67} & \textbf{91.67} & &  \\ \cline{2-8}
\hline

\multirow{6}{*}{USF-MNPAD-I (Heart Rate)} 

& \multirow{3}{*}{Zamzmi et al. \cite{7900284}} & No-pain & 72.58 & \textbf{78.95} & \textbf{75.63} & \multirow{3}{*}{60.82} & \multirow{3}{*}{0.49} \\ \cline{3-6}
& & Pain & 25.00  & 19.05 & 21.62 &  & \\ \cline{3-6}
& & Overall & 59.77 & 62.82 & 61.09 & &    \\ \cline{2-8}

& \multirow{3}{*}{Proposed} & No-pain & \textbf{75.44} & 75.44 & 75.44 &\multirow{3}{*}{\textbf{64.10}} & \multirow{3}{*}{\textbf{0.54}} \\ \cline{3-6}
& & Pain & \textbf{33.33} & \textbf{33.33} & \textbf{33.33} &  &  \\ \cline{3-6}
& & Overall & \textbf{64.10} & \textbf{64.10} & \textbf{64.10}  & &  \\ \cline{2-8}
\hline

\multirow{6}{*}{USF-MNPAD-I (Respiration Rate)} 

& \multirow{3}{*}{Zamzmi et al. \cite{7900284}} & No-pain & \textbf{82.26} & \textbf{89.47} & \textbf{85.71} & \multirow{3}{*}{\textbf{78.73}} & \multirow{3}{*}{\textbf{0.69}} \\ \cline{3-6}
& & Pain & \textbf{62.50} & \textbf{47.62} & \textbf{54.05} &  &  \\ \cline{3-6}
& & Overall & \textbf{76.94} & \textbf{78.20} & \textbf{77.19} & &   \\ \cline{2-8}

& \multirow{3}{*}{Proposed} & No-pain & 77.36 & 71.93 & 74.55 & \multirow{3}{*}{64.10} & \multirow{3}{*}{0.57}\\ \cline{3-6}
& & Pain & 36.00 & 42.86 & 39.13 &  &  \\ \cline{3-6}
& & Overall & 66.22 & 64.10 & 65.01  & & \\ \cline{2-8}
\hline

\end{tabular}
\end{center}
\end{table*}

\subsection{Experimental Details}

\subsubsection{Experiment on MIMIC-III}
This experiment aims to predict an imminent hypotensive episode using Mean Arterial Pressure data obtained from the MIMIC-III dataset. It has been demonstrated by \cite{MarkLast,MOGHADAM2021104120} that we can predict an imminent hypotensive episode by learning patterns from Blood Pressure. As MIMIC-III has sufficient data points to learn patterns, this dataset has been chosen for the experiment.  Adult patients with age range (17-90 years) are considered for this experiment. 
We divide our experiment into 3 parts: Observation window, Gap window, and Target window. We train our model on the data points present in the ``Observation window" which has a duration of 2 hours. Then, we predict if a hypotensive episode will occur in the target window with a gap window of 1 hour. According to \cite{MarkLast}, the hypotensive episode is defined as mean arterial pressure is less than 60 mm Hg for at least 30 minutes. This is the ground truth for our research.

In short, for the experiment, the final dataset includes 142 unique patients (72 hypotensive, 70 non-hypotensive), 1 sample per patient, and a sampling frequency of 1 data-point per hour. The data augmentation has been done 14 times for samples of both the classes by adding random Gaussian noise of Mean = 0 and STD = 3. Finally, total number of samples are 2130  = 70 (no hypotensive original) + 72 (hypotensive original) + 70*14 (no-hypotensive augmented) + 72*14 (hypotensive augmented). 
During the training, VGG16 CNN models are trained for 12 epochs by taking a snapshot of weights for every 2 epochs and final predictions are taken using majority voting from snapshots of 12, 10, 8 epochs. RMSProp optimizer has been used with a learning rate of 0.0001. The number of epochs is less because the validation loss starts increasing after 12 epochs. The model is validated using subject-wise 10-fold cross-validation.

\subsubsection{Experiment on PIC}
In this experiment, Systolic Blood Pressure values obtained during surgery are used to predict an imminent hypotensive episode before 5 minutes of its occurrence. Data points are split into observation window (20 minutes), gap window (5 minutes), and target window (25 minutes), just like we have used for the MIMIC-III dataset. According to University of Iowa Steady Children’s Hospital, hypotension for neonates is defined as systolic pressure less than 60 mmHg. This threshold has been used to generate the ground truth for hypotensive and non-hypotensive episodes. The systolic pressure signal is split into 3 windows namely Observation window of 20 minutes, Gap window of 5 minutes, and Target window of 20 minutes. Using the data from the observation window, we predict with a gap of 5 minutes whether the target window is going to be a hypotensive episode or not. 

In short, for the experiment, the final dataset includes 300 unique patients (225 hypotensive, 75 non-hypotensive), variable number of samples per patient, and a sampling frequency of 1 data-point per 5 minutes. After performing data augmentation with Gaussian Noise of Mean = 0 and STD = 3, the dataset includes  5094 = 255 (hypotensive original ) + 318 (non hypotensive original) + 255*9 (Hypotensive augmented) + 318*7 (Non-Hypotensive augmented) number of samples. 
During the training, VGG16 CNN models are trained for 24 epochs by taking a snapshot of weights for every 3 epochs and final predictions are taken using majority voting from snapshots of 18, 21, 24 epochs. RMSProp optimizer has been used with a learning rate of 0.00001. The model is validated using subject-wise 10-fold cross-validation.

\subsubsection{Experiment on USF-MNPAD-I}
The data considered for the experiment is a 10 seconds sample with a data point for every 1 second. The data is collected in different states like 1 minute before the operation, during the operation, and up to 5 minutes post-operation. Every state has 10 seconds sample along with ground truth values indicating if the baby is experiencing pain or no-pain. 

In short, for the experiment, the final dataset includes 12 unique patients of a total 78 number of samples (57 pain, 21 no-pain), variable number of samples per patient, and a sampling frequency of 1 data-point per second. 
After performing data augmentation, the dataset includes  1137 =  57 (no-pain original) + 21 (pain original) + 57*9 (No-pain augmented) + 21*26 (Pain augmented) number of samples. Data augmentation is done by adding Gaussian noise with Mean = 0 and STD = 3. As the number of subjects, samples and consequently the spectrograms are less, VGG16 may not be a good model to train as it is way too complex with millions of parameters and may over-fit. To overcome this problem, a shallow model (Table \ref{model}) is adopted which has fewer parameters. The model shown in Table \ref{model} has 113,489 parameters which are lesser than VGG16 and does not tend to overfit. 
The experiment is done using subject-wise leave-one-out cross-validation and the snapshot ensemble method is used on every fold. During the training, the model was trained for 45 epochs by taking a snapshot of weights for every 3 epochs and final predictions are taken using majority voting from snapshots of 39, 42, 45 epochs. RMSProp optimizer has been used with a learning rate of 0.0001. The model is validated using subject-wise leave-one-out cross-validation.

\begin{table}[t]
\caption{Architecture of the shallow model for USF-MNPAD-I}
\label{model}
\begin{center}
\begin{tabular}{|c|c|}
\hline
\textbf{Layers} & \textbf{Details}\\
\hline
Conv2D & $128\times3\times3$, stride = 1, pad = same\\
\hline
Conv2D & $128\times3\times3$, stride = 1, pad = same\\
\hline
MaxPooling & pool size = $2\times2$\\
\hline
Conv2D & $64\times3\times3$, stride = 1, pad = same\\
\hline
Dense & 1, Sigmoid\\
\hline
Total parameters & 113,489\\
\hline

\end{tabular}
\end{center}
\end{table}

\subsubsection{Experiment on NEBD}
The experiment performed in this dataset is about classifying stress and relaxed states of 20 subjects based on their Heart Rate signal. 
In short, for the experiment, the final dataset includes 20 unique patients of a total 120 number of samples (60 stress, 60 relax), 6 number of samples per patient, and a sampling frequency of 1 data point per second. The dataset has ground truth information regarding stress and relaxation. 
After performing data augmentation, with Gaussian Noise (Mean = 0 and STD = 1) the dataset includes  total 3480 (60*28 augmentations for stress + 60*28 augmentations for relaxation + 60 original for stress + 60 original for relaxation) number of samples. 
During the training, VGG16 models were trained for 30 epochs by taking a snapshot of weights for every 3 epochs and final predictions are taken using majority voting from snapshots of 24, 27, 30 epochs. RMSProp optimizer has been used with a learning rate of 0.00001. The model is validated using subject-wise leave-one-out cross-validation.

\subsection{Result Analysis}
From different experiments (Table \ref{mimic}), it can be observed that the proposed method can perform well on different vital signs signals such as Blood Pressure, Heart Rate, Respiratory Rate while using reconstructed spectrogram compared to regular spectrogram image. 

In case of prediction tasks, for the MIMIC-III dataset \cite{MIMICIII}, the proposed method has better precision, recall, F1-score for hypotensive class providing an accuracy of 91.55\% compared to the baseline method (90.67\%). In case of the PIC dataset \cite{PIC}, to the best of our knowledge, there is no research available that predicts an imminent hypotensive episode for neonates during surgery. The proposed method has better overall precision, recall, F-1 score and providing an accuracy of 87.61\% compared to the baseline method (86.90\%).

For classification tasks, the proposed approach shows both better accuracy and AUC. In case of the NEBD dataset \cite{7780081}, which is about classifying stress and relaxation, the proposed method shows better performance than the baseline method in all the metrics. In case of the USF-MNPAD-I dataset of Heart Rate, the proposed method had better overall precision, recall, F-1 score, accuracy, AUC compared to the existing baseline approach. But in the case of Respiratory Rate, the baseline approach performs better than the proposed method in all the metrics. That reveals that for respiration rate it is not able to generate a good pattern using the spectrogram features.

\section{Conclusion}
Spectrograms are well suited for audio signals to capture frequency variations. However, in the case of vital signs, it is difficult to use regular spectrograms as input to CNN as they do not show patterns and variations like audio signals. To solve this issue, in this paper, we proposed applying frequency modulation on vital signs and applying spectrograms on FM signals. The proposed method shows better performance in terms of several metrics on several datasets. This method can be helpful when using multi-modal approach when we want to combine results from audio, video and time series data using CNNs.  We can also expand the usage of the proposed method to other uni-variate time-series signals such as temperature, EDA, accelerometer readings.






\bibliographystyle{IEEEtran}
\bibliography{ref}

\end{document}